\documentclass[aps,prd,superscriptaddress,amsmath,amssymb,showpacs,twocolumn]{revtex4-2}
\usepackage[cp866]{inputenc}
\usepackage{amssymb,bm}
\usepackage{graphicx}
\usepackage{amsmath}
\usepackage{color}
\usepackage[colorlinks=true,citecolor=blue,urlcolor=blue]{hyperref}
\begin{document}

	\title{Invariant-mass spectrum of $\Lambda\bar\Lambda$ pair in the process $e^+e^-\rightarrow\phi\Lambda\bar\Lambda$}

	\author{A.I. Milstein}
	\email{A.I.Milstein@inp.nsk.su}
	\author{S.G. Salnikov}
	\email{S.G.Salnikov@inp.nsk.su}
	\affiliation{Budker Institute of Nuclear Physics of SB RAS, 630090 Novosibirsk, Russia}
	\affiliation{Novosibirsk State University, 630090 Novosibirsk, Russia}
	\date{\today}

\begin{abstract}
We show that the final-state interaction perfectly explains the recent experimental data of BESIII Collaboration on the near-threshold invariant-mass spectrum of $\Lambda\bar\Lambda$ in  the process $e^+e^-\rightarrow\phi\Lambda\bar\Lambda$. The position of peak in invariant-mass spectrum is above the threshold though the interaction of $\Lambda$ and $\bar\Lambda$ is due to an attractive potential. This potential has a simple form and depends  only on two parameters. We show that the invariant-mass spectrum is consistent with the quantum numbers $J^{PC}=1^{++}$ and $2^{++}$  of $\Lambda\bar\Lambda$ pair but contradicts to $J^{PC}=2^{-+}$.
\end{abstract}
	\maketitle

Nowadays, many resonances have been observed with masses only slightly above the threshold of the production of particles into which these resonances mainly decay. Although there are a lot of experimental data, the discussion about the nature of these resonances still continues. Very recently,
the BESIII Collaboration has published the experimental results \cite{BES32021} on annihilation $e^+e^-\rightarrow\phi\Lambda\bar\Lambda$.  These results clearly demonstrate a strong enhancement of  $\Lambda\bar\Lambda$ pair production in the near-threshold  mass region. A study of the angular distributions carried out in Ref.~\cite{BES32021} shows that the possible quantum numbers of  $\Lambda\bar\Lambda$  pair are 
$J^{PC}=1^{++},\,2^{++},\,2^{-+}$. In the first two cases, the relative angular momentum $l$ of $\Lambda\bar\Lambda$ pair, its total spin $S$, and the relative angular momentum $L$ of a pair and $\phi$ meson are $l=1$, $S=1$, and $L=0$. In the third  case, the corresponding quantum numbers are   $l=2$, $S=0$, and $L=1$.

The invariant-mass spectrum of  $\Lambda\bar\Lambda$ is similar to that observed in $\Upsilon (4S) \rightarrow B^0\bar B^0$ and  $\Upsilon (4S) \rightarrow B^+B^-$ decays \cite{BABAR2009, Dong2020,Belle2021}.
These decays were discussed within various theoretical  approaches \cite{Marciano1990, Lepage1990,Eichten1990,Kaizer2003,Voloshin2003,Voloshin2005,Oset2020} with completely different qualitative predictions. In our recent work \cite{MS2021} a simple exactly solvable model has been proposed for describing   $\Upsilon (4S) \rightarrow B^0\bar B^0$  and  $\Upsilon (4S) \rightarrow B^+B^-$ decays. This model is based on the account for the final-state interaction of $B$ and $\bar B$ mesons. The results of \cite{MS2021} are in good agreement with the available experimental data. The method used in \cite{MS2021} is similar to that  developed earlier in \cite{DMS2014, DMS2016, MS2018} for calculating the cross section of  $e^+e^-$ annihilation into  proton-antiproton  and  neutron-antineutron pairs   near the pair production threshold.
The effects of final-state interaction in the hadronic systems were also discussed in \cite{Mei2005,Mei2006,Mei2015,Mei2018}.

The approach of \cite{MS2021} can also be applied for calculation of $\Lambda\bar\Lambda$ invariant-mass spectrum. In fact, in our case it is even simpler because there is no Coulomb interaction and there is only isoscalar exchange between $\Lambda$ and $\bar\Lambda$.   Of course, at present it is not known much about the corresponding potential $V(r)$ of $\Lambda\bar\Lambda$ interaction. However, for our purposes we can use any simple phenomenological potential with some specific properties (see below). Our approach ensures agreement between the theoretical predictions and the experimental data. Besides,  it clarifies the physics of near-threshold resonances. This is the main goal of our work.
  
Following \cite{MS2021}, consider the radial wave function $\Psi(r)$ of  $\Lambda\bar \Lambda$ pair. The function $\psi(r)=kr\,\Psi(r)$ satisfies the equation
\begin{align}\label{we}
& \left[-\dfrac{1}{M}\,\dfrac{\partial^2}{\partial\,r^2}+
\dfrac{l(l+1)}{Mr^2}+V(r)-E \right]\psi(r)=0\,,
\end{align}
where $M$ is the mass of $\Lambda$, $k=\sqrt{ME}$, and $E$ is the energy of the pair, counted from $M_{th}=2M$, $\hbar=c=1$. 
It is necessary to find a solutions $\psi(r)$ of \eqref{we} with asymptotics at large distances
\begin{align}
& \psi(r)=\frac{1}{2i}\left(S\,\chi^{+}-\chi^{-}\right)\,,\nonumber\\
&\chi^{\pm}=\exp\left[\vphantom{\bigl(\bigr)}\pm i\left(kr- l\pi/2\right)\right],
\end{align}
where $S$ is some function of energy. The factors $W_l$, 
 \begin{align}\label{3W}
&W_1=N_1\,k\,\left|\dfrac{\partial}{\partial r}\,\Psi(0)\right|^2\quad \mbox{for}\quad l=1\,,\nonumber\\
&W_2=N_2\,k\,\left|\dfrac{\partial^2}{\partial r^2}\Psi(0)\right|^2\quad \mbox{for}\quad l=2\,,
\end{align}
where $N_{1,2}$ are some constants, determine a strong  energy dependence of $\Lambda\bar\Lambda$ invariant-mass spectrum. There is also smooth dependence on $E$ in the cross section of annihilation $e^+e^-\rightarrow\phi\Lambda\bar\Lambda$ via the momentum of $\phi$ meson. However, we may neglect this dependence in the vicinity of $\Lambda\bar\Lambda$ pair production threshold.

 As a model, we choose a simplest potential $V(r)=-V_0\,\theta(a-r)$, where $ V_0>0 $ and $a$ are some parameters, $\theta(x)$ is the Heaviside function. These parameters can be chosen so that the corresponding results describe well the available experimental data. For this potential, the analytical solution has a simple form  \cite{MS2021},
 \begin{align}\label{w12}
 	&W_1=b_1\, \dfrac{k\,q^2}{M^3}\,\Bigg|\dfrac{q\,}{k\,h_1'(ka)\,f_1(qa) -q\, h_1(ka)\,f_1'(qa)}\Bigg|^2,\nonumber\\
 	&W_2=b_2\, \dfrac{k\,q^4}{M^5}\,\Bigg|\dfrac{q\,}{k\,h_2'(ka)\,f_2(qa) -q\, h_2(ka)\,f_2'(qa)}\Bigg|^2,\nonumber\\
 	& h_1(x)=\left(\dfrac{1}{x}- i\right)\,e^{ix}\,,\quad f_1(x)=\dfrac{\sin x}{x}-\cos x\,,\nonumber\\
 	&h_2(x)=\left(\dfrac{3}{x^2}-\dfrac{3i}{x}-1\right)\,e^{ix}\,,\nonumber\\ &f_2(x)=\left(\dfrac{3}{x^2}- 1\right)\sin x-3\dfrac{\cos x}{x}\,,
 \end{align}
 where $q=\sqrt{M(E+V_0)}$, $b_{1,2}$ are some energy-independent constants, and $Z'(x)\equiv \partial Z(x)/\partial x$. As should be, $W_1 \propto k^3$ and $W_2 \propto k^5$ at $E\to0$. However, these asymptotics are valid only for very small energy $E$.

\begin{figure}[b]
	\centering
	\includegraphics[width=\linewidth]{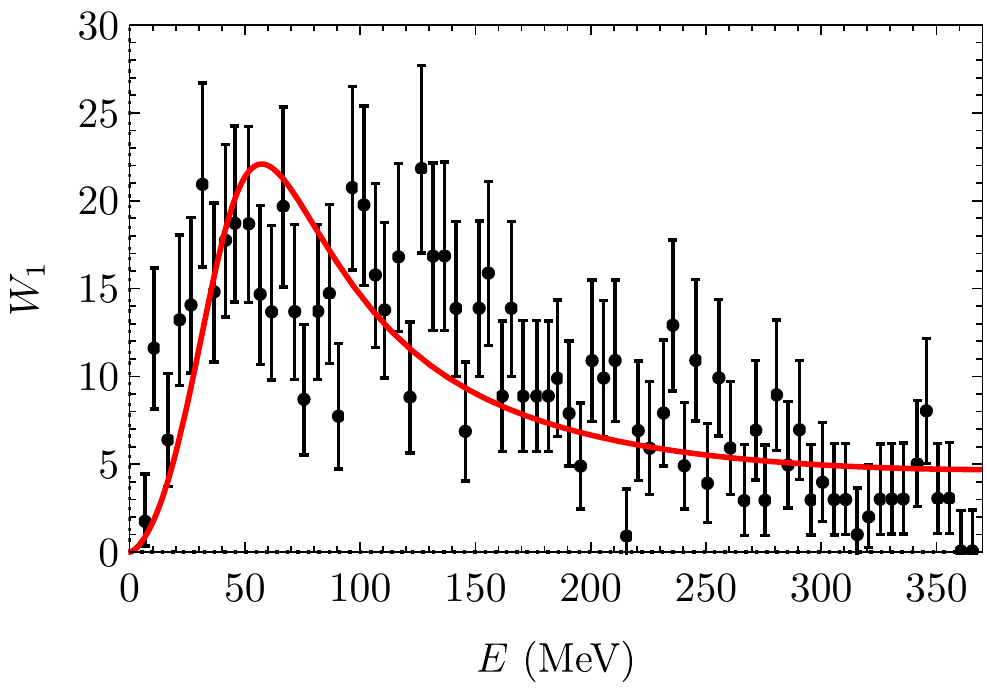}
	\caption{Energy dependence of the function $W_1$ (solid curve) normalized to the experimental data (events/($5~\mathrm{MeV}$)) at $a=1.2\,\mbox{fm}$ and $V_0=836\,\mbox{MeV}$. The experimental data are from Ref. \cite{BES32021} (BESIII).}
	\label{FigLamLam}
\end{figure}

The observed peak in the invariant-mass spectrum appears due to a strong reflection from the border of the potential, so that $q_Ra=\sqrt{M(V_0+E_R)}\,a=n\pi$, where $n=1,2,\dots$ and $E_R$ is a parameter close to the value of the peak position, $E_R\approx 100 \,\mbox {MeV}$. Thus,
\begin{equation}\label{V0}
V_0=\dfrac{(n\pi)^2}{Ma^2} -E_R\,.
\end{equation}
In the case of $e^+e^-\rightarrow N\bar N$ annihilation, a nucleon-antinucleon pair is produced in the superposition of $l=0$ and $l=2$ waves. As a result, it is necessary to use an optical potential with the negative imaginary part, which accounts for annihilation of $N\bar N$ pair into mesons. In our cases $l=1$ or $l=2$, hence annihilation of $\Lambda\bar\Lambda$ into mesons is not important, so that we use a real value of $V_0$.  The dependence of the function $W_1$ on $E$ \eqref{w12} is shown in
Fig. \ref{FigLamLam} for $a=1.2\,\mbox{fm}$ and $V_0=836\,\mbox{MeV}$, data are taken from Ref. \cite{BES32021} (BESIII). Values  $V_0$, $a$ of the potential and the external factor $b_1$ in $W_1$ minimize $\chi^2$. We obtain  $\chi^2/N_{df}=1.2$, and this number indicates good agreement between our predictions and the experimental data. The values of the parameters $a$ and $V_0$ correspond to $n=2$ and $E_R=106\,\mbox{MeV}$, see Eq.~\eqref{V0}. The relation $E_R \ll V_0$ corresponds to strong reflection from the border of the potential.

Note that the experimental data shown in Fig.~\ref{FigLamLam} were obtained in Ref.~\cite{BES32021} by summing up the data samples taken at different values of $e^+$ and $e^-$ beam energies. However, we have checked that our predictions (scaled to the number of events) are in good agreement with the data from each sample given in supplementary materials to Ref.~\cite{BES32021}. Therefore, strong dependence of the cross section of  $e^+e^-\to\phi\Lambda\bar\Lambda$ annihilation on the invariant mass of $\Lambda\bar\Lambda$ pair close to the threshold is given solely by the functions $W_l$.

In the nonrelativistic approximation, spin-orbit interaction is not taken into account and  the function $W_1$ is the same for $J^{PC}=1^{++}$ and $J^{PC}=2^{++}$ states, since in both cases $l=1$. 
 In the case $l=2$, our attempts to achieve a small $\chi^2$ for reasonable values of $V_0$ and $a$  were not successful. This allows us to claim that the case $l = 2$ ($J^{PC}=2^{-+}$) is inconsistent with the experimental data.
 
To conclude, it is shown that our predictions for the invariant-mass spectrum of $\Lambda\bar\Lambda$ in  $e^+e^-\rightarrow\phi\Lambda\bar\Lambda$ annihilation are in good agreement with the recent experimental data of BESIII Collaboration.  These predictions are based on the account for the final-state interaction within a simple model. We claim that $\Lambda\bar\Lambda$ pairs are produced mainly in the states with quantum numbers $J^{PC}=1^{++}$ or $2^{++}$ but not with $J^{PC}=2^{-+}$. Our results for $\Lambda\bar\Lambda$ invariant-mass spectrum strongly support the statement that the origin of near-threshold peaks in various processes is related to the final-state interaction of the produced hadrons.

 We are grateful to A.E.~Bondar for useful discussions.


\begin{thebibliography}{99}
\bibitem{BES32021}	 M. Ablikim {\it et. al.} [BESIII Collaboration], Phys. Rev. {\bf D 104}, 052006 (2021).
\bibitem{BABAR2009} B.Aubert {\it et. al.} [BABAR Collaboraion], Phys. Rev. Lett. {\bf 102}, 012001 (2009).	
\bibitem{Dong2020} Xiang-Kun Dong, Xiao-Hu Mo, Ping Wang, Chang-Zheng Yuan, Chinese Phys. {\bf C44}, 083001 (2020). 
\bibitem{Belle2021} R.Mizuk {\it et. al.} [Belle Collaboraion], arXiv: 2104.08371 [hep-ex] (2021).	
\bibitem{Marciano1990} D. Atwood and W.J. Marciano, Phys. Rev. {\bf D41},  1736 (1990).
\bibitem{Lepage1990} G.P. Lepage, Phys. Rev. {\bf D42}, 3251 (1990).	
\bibitem{Eichten1990} N. Byers and E. Eichten, Phys. Rev. {\bf D42},  3885 (1990).
\bibitem{Kaizer2003} R. Kaiser, A.V. Manohar, and T. Mehen, Phys. Rev. Lett. {\bf 90}, 142001 (2003).
\bibitem{Voloshin2003} M.B.Voloshin, Mod. Phys. Lett. {\bf A18}, 1783 (2003). 
\bibitem{Voloshin2005} M.B.Voloshin, Yad. Fiz. {\bf 68}, 804 (2005) [Phys.   At. Nucl. {\bf 68},  771 (2005)].
\bibitem{Oset2020} W. H. Liang, N. Ikeno, and E. Oset, Phys. Lett. {\bf B 803},
135340 (2020).
\bibitem{MS2021} A.I. Milstein and S.G. Salnikov, Phys. Rev. {\bf D 104}, 014007 (2021).
\bibitem{DMS2014}  V. F. Dmitriev, A. I. Milstein, S. G. Salnikov,
Yad. Fiz. {\bf 77}, 1234 (2014) [Phys. At. Nucl. {\bf 77},  1173 (2014)].
\bibitem{DMS2016}  V.F. Dmitriev, A.I. Milstein, and S.G. Salnikov,
Phys. Rev. {\bf D 93}, 034033 (2016).
\bibitem{MS2018}  A.I. Milstein,  S.G. Salnikov,
	Nucl. Phys. {\bf A 977}, 60  (2018).
\bibitem{Mei2005} A. Sibirtsev, J. Haidenbauer, S. Krewald, U.-G.
Mei{\ss}ner, and A. W. Thomas, Phys. Rev. {\bf D 71}, 054010
(2005).
\bibitem{Mei2006} J. Haidenbauer, U.-G. Mei{\ss}ner, and A. Sibirtsev, Phys. Rev.
{\bf D 74}, 017501 (2006).
\bibitem{Mei2015} X.-W. Kang, J. Haidenbauer, and U.-G. Mei{\ss}ner, Phys. Rev.
{\bf D 91}, 074003 (2015).	
\bibitem{Mei2018}Ling-Yun Dai, Johann Haidenbauer, and U.-G. Mei{\ss}ner, Phys. Rev. {\bf D 98}, 014005 (2018).
	
\end{thebibliography}
 \end{document}